\documentclass{aa}
\usepackage{graphicx}
\begin{document}
 
\title{Properties of disks and spiral arms
along the Hubble sequence}
\author{Jun Ma}
 
\offprints{Jun Ma, \\
\email{majun@vega.bac.pku.edu.cn}}
\institute{National Astronomical Observatories,
Chinese Academy of Sciences, Beijing, 100012, P.R. China}
\date{Received........; accepted........}
 
\abstract{
This paper presents
some new statistical correlations between the properties of
disks and spiral arms with some physical properties
of galaxies. Our results show that
the thickness of spiral disks tend to diminish along
the Hubble sequence in the sense that disks of Sc galaxies are
40\% thinner than disks of Sab-Sb galaxies. Moreover,
the thinner disks tend also to be bluer.
We also find that there exists a correlation between HI linewidths
and arm pattern within each Hubble type, which suggests that the arm
shape is partially determined by the mass of a galaxy.
Total mass luminosity ratios and total mass surface densities
also have correlations with pitch angles, i.e., for disks
with lower surface densities and lower total mass luminosity
ratios, the pitch angles tend to be greater.
} 

\titlerunning{Properties of disks and spiral arms
along the Hubble sequence}
\authorrunning{J. Ma}
\maketitle

\section{Introduction}

When investigating the mass distribution within highly flattened,
axisymmetric, self-gravitating spiral galaxies, Toomre (1963)
found a family of exact solutions of the Poisson equation
with arbitrary laws of rotation,
\begin{equation}
\bigtriangledown^2\phi(R,z)=-4\pi G \rho(R,z)=-4\pi G\mu(R)\delta(z).
\end{equation}
From Toomre's (1963) disk models of galaxies, Miyamoto \&
Nagai (1975) introduced a new potential-density pair,
in which the mass is no longer confined to an infinitesimally
thin disk. This model is free from singularities and involve
functions that are quite explicit and elementary.

Lin \& Shu (1964, 1966) did the pioneering work to explain
the spiral structure of disk galaxies using density
waves to describe spiral patterns. These waves compress
the gas component as it flows
through the arms (Roberts, Roberts \& Shu, 1975).
The spiral arms can also be the result of the tidal field of
a nearby neighbor (Toomre \& Toomre, 1972), bars
(Sanders \& Huntley, 1976) or Stochastic
Self-propagating Star Formation (Gerola \& Seiden, 1978;
Mueller \& Arnett, 1976).

Under the {\it quasi-stationary spiral structure theory}
(QSSS) of density waves by Lin \& Shu (1964) and
using the Green function method plus Fourier-Bessel transformation
of three dimensional disk galaxies, Peng et al. (1978, 1979,
1988) obtained the formulation of the gravitational potential
of both galactic disks and spiral arms with different
mathematical spiral curves. They generalized the Toomre
models of two dimensional disk galaxies to three
dimensional models. That is to say, the distribution
of surface density may be analytically derived from
Toomre's rotation curves in a disk galaxy with
a finite thickness. Based on the fundamental assumption
(see details in Peng et al. 1979 or Peng 1988)
that the density distribution along
$z$-direction for a finite thickness disk is
\begin{equation}
\rho(r, \phi, z)=\frac{1}{H}\sigma(r, \phi)e^{-|z|/h},
\end{equation}
where $h$ is the vertical exponential scale height,
%. In this expression $h$
%represents the vertical exponential scale height,
$H=2h$ the disk
thickness and $\sigma(r, \phi)$ is
the surface density, Peng et al. (1979)
obtained a criterion for the appearance of density waves, which is
\begin{equation}
r>r_0=\frac{H\sqrt{m^2+\Lambda^2}}{2},
\end{equation}
where $r_0$ is the polar coordinate of the
starting point from which the arms of a galaxy stretch outward on
the galactic plane, $m$ is the number of the arms in a
spiral galaxy, and $\Lambda$ is the winding parameter of the arm.
Based on this criterion, Peng (1988) proposed
a method for estimating the thickness of a non-edge-on spiral
galaxy. Using this method, Ma et al. (1997, 1998) estimated the thickness of
72 spirals.

To study the component separation in edge-on
spiral galaxies, van der Kruit \& Searle (1981a,
1981b, 1982a, 1982b) proposed a model for the
three-dimensional distribution of light in galactic
disks. Under the assumption that a disk galaxy
has a locally isothermal, self-gravitating and
truncated exponential disk, the space-luminosity of
this model can be described by
\begin{equation}
L(r, z)=L_0e^{-r/h_r}{\rm{sech}}^{2}(z/z_0).
\end{equation}
A detailed investigation of  edge-on
spiral galaxies by de Grijs
et al. (1997) showed, however, that  the vertical light profiles
are much closer to the exponential than to the isothermal
solution.

The question of the mathematical form of spiral arms was recognized at the
beginning of this century (von der Pahlen, 1911; Groot, 1925). Then,
Danver (1942), Kennicutt (1981) and Kennicutt \& Hodge (1982)
systematically studied the shapes of
spiral arms by fitting different spiral curves by least squares and
using an iterative procedure. These authors concluded that spiral
arms are better represented by logarithmic spirals.

There are many determinations of the pitch angles of spiral
arms that can be found in the literature.
Danver (1942) measured these values for a sample of
98 nearby spirals.
Kennicutt (1981) measured the shapes of spiral arms
in 113 nearby Sa-Sc galaxies.
Using the Fourier analysis first proposed by Kalnajs (1975)
and further developed by Consid\`ere \& Athanassoula (1982, 1988),
Garc\'{\i}a-G\'omez \& Athanassoula (1993) calculated
the pitch angles in a sample of 44 spiral galaxies.
Puerari \& Dottori (1992) using also a similar technique
measured the pitch angles in 22 galaxies. Recently,
Seigar \& James (1998a, 1998b) determined
the pitch angles of spiral arms for 45 spiral galaxies based on the
images in the near-IR J and K bands.
With the IRAF software,
Ma et al. (1997, 1998) directly fitted the shapes of
spiral arms on the images of 72 northern spiral galaxies.
The details of this procedure can be found in Ma et al. (1997, 1998, 1999).
These measures on the pitch angles can be used to understand which is the
influence of the different physical properties of galaxy disks in
arm shapes.

The structure of this paper is as follows: In Sect. 2 we present
a correlation between galaxy type or color with disk thickness. In
Sect. 3, we show some new correlations between the pitch angles
of spiral arms and the physical properties of galaxy disks. Finally,
in Sect. 4 we present our conclusions.

\section{Statistical properties of disk thickness along the Hubble sequence}

Our statistical sample contains 72 northern spiral galaxies, for which the
values of disk thickness, inclination and pitch angle of individual
spiral arms were measured in  Ma et al. (1997, 1998).
A more detailed discussion can be found in Ma et al. (1999).

Ma et al. (1998, 1999) have presented some
statistical correlations between thickness and flatness of
spiral disks with some other physical properties of spiral
galaxies. The main conclusions are that the thickness is
correlated with color, mass in neutral hydrogen and 
$\rm {H}_{\alpha}+[\rm {NII}]$ equivalent width
(see details in Ma et al. 1998, 1999).
In this section, we will investigate a new
correlation between thickness or flatness and total
luminosity.

\begin{figure}
\resizebox{\hsize}{!}{\rotatebox{-90}{\includegraphics{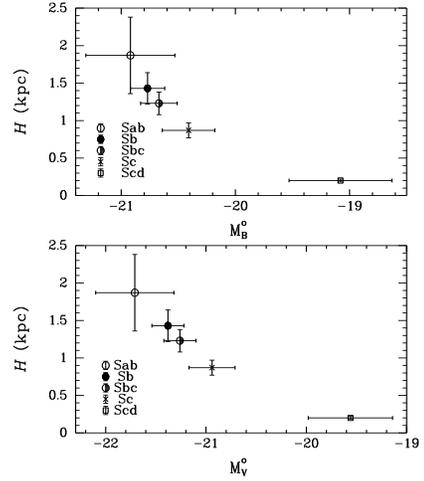}}}
\caption{Mean thickness of spiral galaxy disks plotted against the mean
total (``face-on'') absolute magnitude in the B and V systems
for each Hubble type of the galaxies in our sample. The upper
panel shows this relation for the B magnitudes while the lower
panel shows this relation for the V magnitudes.}
\label{FigVibStab}
\end{figure}

Roberts \& Haynes (1994) systematically studied the physical parameters
along the Hubble sequence by making use of two primary catalogues, the first one
is the RC3 catalogue (de Vaucouleurs et al., 1991) and the second
one is a private catalogue maintained by R. Giovanelli \&
M. Haynes. These authors showed that the mean total luminosity of
the galaxies changes only slightly as a function of Hubble type
until the latest types (Scd). The mean total luminosity,
however, clearly shows a decrease for the dwarf galaxies (Sd and Sm).

In order to study the correlation
between thickness along the Hubble sequence and total
(``face-on'') absolute magnitude from the RC3
catalogue (de Vaucouleurs et al., 1991),
we plot in Fig. 1 the mean thickness for each Hubble type of
the galaxies in our sample versus mean total luminosity.
In this figure a clear trend is shown,
namely, galaxies of later Hubble types tend to be flatter
than galaxies of early types. More specifically, Scd galaxies are,
on average, 40\% flatter than Sb galaxies.
The total absolute optical magnitudes in the B and V systems,
which are corrected for differential galactic and internal extinction
(to ``face-on''), and the Hubble types, come from the RC3
catalogue (de Vaucouleurs et al., 1991).
We calculated the mean, dispersion of the distributions of $H$
between different Hubble types.
%and total absolute optical magnitude in the B system between
%different Hubble types.
Table 1 lists the relevant
quantities for these distributions. In Col. 1 we give the Hubble
type and in Col. 3 the numbers of galaxies in each Hubble type,
as well as the difference. Col. 4 gives the means of the distributions
of $H$, Col. 5 the dispersion around
the means. In Col. 6 we present the probability that the
two $H$ means
between different Hubble types come from the same distribution.
Since there are only 4 Sab galaxies in this study, we merge
these galaxies with the Sb galaxies to get a single Sab-Sb class.
From the Student's t-test we can see that
the means of the distributions of $H$
between different Hubble types are different
except for the means between Sab-Sb class and Sc class.

Guthrie (1992) derived the axial ratios of disk components, $R$, for
262 edge-on spiral galaxies, and analyzed the distribution of
isophotal axial ratios for 888 diameter-limited normal Sa-Sc
galaxies to obtain information
on the true axial ratios $R_0$. If we suppose that the
radial to vertical scale parameter ratio is equal
to $R_0$ from Guthrie (1992), we can calculate
the values of $\overline{\log R_0}$ for each Hubble
type of the galaxies in our sample.
 
Fig. 2 shows the dependence of $\overline{\log R_0}$ on galaxy type
in the RC3 catalogue (de Vaucouleurs et al., 1991).
In this figure we can see a correlation between $\overline{\log R_0}$
and Hubble sequence, in the sense that galaxies become systematically
flatter when going from type Sab to Scd. We also calculated
the mean, dispersion of the distribution of $\overline{\log R_0}$
between different Hubble types. Table 2 summarises the relevant
quantities of this distribution as in Table 1.
We also merge
Sab galaxies with the Sb galaxies to get a single Sab-Sb class.
From the Student's t-test we can see that
the means of the distributions of $\overline{\log R_0}$
between different Hubble types
are also different
except for the means between
Sab-Sb class and Sbc class.

\begin{figure}
\resizebox{\hsize}{!}{\rotatebox{-90}{\includegraphics{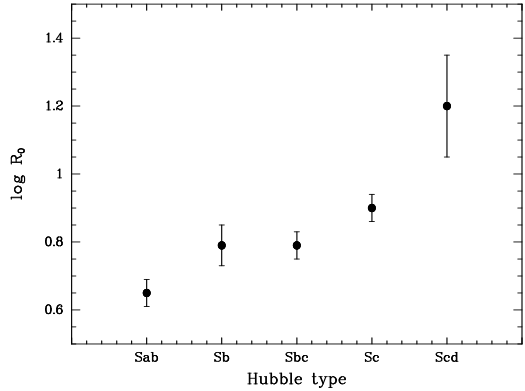}}}
%\vspace{-1cm}
\caption{Dependence of $\overline{\log R_0}$ on galaxy type in RC3}
\end{figure}

Ma et al. (1998) have studied the correlation between thickness and
color index ($(B-V)_{T}^{0}$) between the B and V bands,
and presented the conclusion that thinner
spirals are bluer. Fig. 3 plots the correlation of average thickness
with color index ${(B-V)}_{T}^{0}$ along the Hubble type in the RC3
catalogue (de Vaucouleurs et al., 1991).
The color indices, which are corrected for differential
galactic, internal extinction (to ``face-on'') and
redshift between the B and V bands, are also taken from the RC3
catalogue (de Vaucouleurs et al., 1991).
The corresponding averages for each Hubble type
are showed in Fig. 3 as in Fig. 1.
Fig. 3 shows that the correlation between
average thickness and color index is also monotonic.
Namely, along the Hubble sequence the galaxies tend to
be bluer and thinner. We also calculated
the mean, dispersion of the distribution of ${(B-V)}_{T}^{0}$
between different Hubble types. Table 3 summarises some of the relevant
quantities for this distribution as in Tables 1 and 2.
We also merge
Sab galaxies with the Sb galaxies to get a single Sab-Sb class.
From the Student's t-test we can see that
the means of the distributions of ${(B-V)}_{T}^{0}$
between different Hubble types
are not different except for 
the means between
Sc class and Scd class.

\begin{figure}
\resizebox{\hsize}{!}{\rotatebox{-90}{\includegraphics{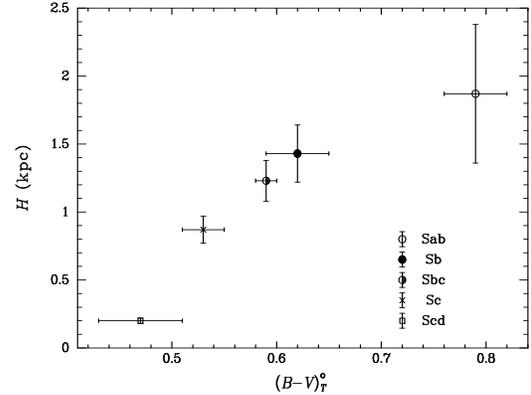}}}
%\vspace{-1cm}
\caption{Thickness of spiral galaxy disks plotted against the
corrected $B-V$ color. Both thickness and color are the corresponding
means for each Hubble type of the galaxies in our sample.}
\label{FigVibStab}
\end{figure} 

\section{Statistical properties of spiral arms with some physical properties of galaxies}

\subsection{Comparison of our pitch angle values with other sources}

In this section, we will show some new correlations
between the properties of spiral arms
and some physical properties of galaxy disks.
But, first, it will be useful to compare
the pitch angles of the common galaxies,
which were measured by Danver (1942), Kennicutt (1981) or
other authors (Consid\`ere \& Athanassoula, 1988;
Puerari \& Dottori, 1992; Garc\'{\i}a-G\'omez \& Athanassoula,
1993). The comparisons are plotted in Fig. 4. In this
figure, the dashed lines are the diagonal, and
the solid lines are the result of the linear correlations.
For each line, we calculated the correlation coefficient,
which is shown in Table 4, together with the number of
points used for each correlations. In Col. 1
we list the symbol of panel in Fig. 4 and in Col. 2
the number of galaxies. Col. 3 gives the
linear correlation coefficient.
From Table 4 and Fig. 4, we can see that our data
correlate better with the values of Danver and Kennicutt
than the data coming from the rest of the sources.
The reason may be that our method is similar to the method
used by Danver and Kennicutt.

\begin{figure}
\resizebox{\hsize}{!}{\rotatebox{-90}{\includegraphics{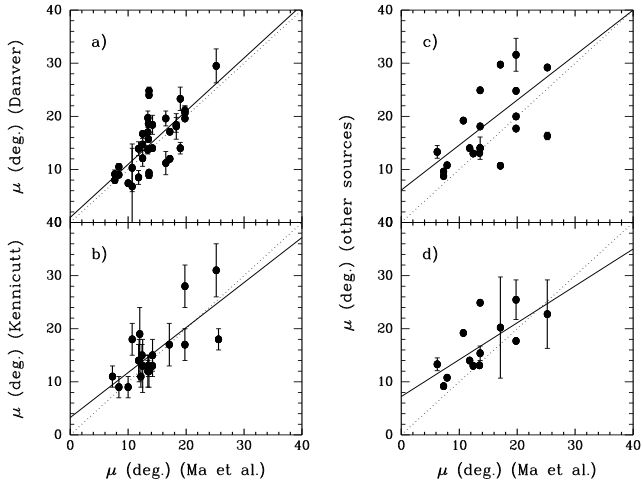}}}
\caption{Comparison of our pitch angle results (Ma et al.,
1997, 1998) with a) Danver (1942),
b) Kennicutt (1981), and c) Consid\`ere
\& Athanassoula (1988), Puerari \& Dottori (1992), and
Garc\'{\i}a-G\'omez \& Athanassoula (1993),
d) mean pitch angles from the sources of c),
for the galaxies in common. In the panel of d), the vertical
error bars are the r.m.s. calculated from
all values in the sources of c).}
\end{figure}

\subsection{Pitch angles as a function of Hubble type}
 
Hubble (1926, 1936) introduced an scheme to classify galaxies
which is still in use today.
Galaxy types run from elliptical,
to lenticular, spiral and finally, irregular galaxies.
This scheme, which was extended by some astronomers
(Holmberg, 1958; de Vaucouleurs, 1956, 1959; Morgan, 1958, 1959;
van den Bergh, 1960a, b, 1976; Sandage, 1961;
Sandage \& Tammann, 1981, 1987; Sandage \& Bedke, 1993),
is based on a number of observational criteria such as
the gas content, size of nuclear bulges,
resolution of the spiral arms, and pitch angle of
spiral arms.
 
The tightness of spiral arms, in addition to the degree of resolution
in the arm and
the relative size of the unresolved nuclear region, are the fundamental
criteria in
Hubble's classification of spiral galaxies (Hubble, 1926, 1936). In order to better
understand the nature and origin of the Hubble sequence and evaluate the
difference in the classification, Kennicutt (1981) compared his measures of pitch angles
with the Hubble type as determined by Sandage \& Tammann
(1981, hereafter ST) and Yerkes class by Morgan (1958, 1959).
From Figs. 7 and 8 in Kennicutt (1981), we can see that, although
the ST's classification is
based almost solely upon disk resolution and the Morgan's solely on the central
concentration of the galaxies,
the trends between arm pitch angles and the galaxy types in
both classifications are almost
the same. In the classifications of galaxies by de Vaucouleurs et al.
(1976, RC2; 1991, RC3), the three criteria are considered.
 
\begin{figure}
\resizebox{\hsize}{!}{\rotatebox{-90}{\includegraphics{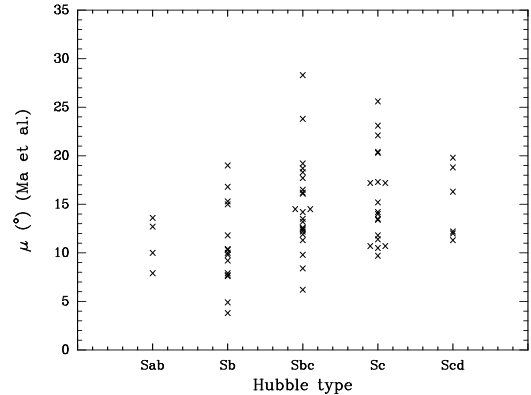}}}
%\vspace{-1cm}
\caption{Mean measured pitch angle plotted against different Hubble types in RC3.}
\end{figure}
 
To clearly see the weight of the tightness
of spiral patterns in the RC3 catalogue (de Vaucouleurs et al., 1991),
we plot the measured pitch angles by
Ma et al. (1997, 1998) against the Hubble types coming
from the RC3 catalogue (de Vaucouleurs et al., 1991) in Fig. 5.
We can see that there is an
increase of the mean pitch angles along the Hubble sequence.
This trend represents the spirit of Hubble's
original scheme for the classification of spirals.
This relation has been discussed by many authors such as
Kennicutt (1981), Consid\`ere \& Athanassoula (1988),
Puerari \& Dottori (1992), and Garc\'{\i}a-G\'omez \& Athanassoula (1993).
Our measures on the pitch angles follow the expected relation with
Hubble type as for the rest of authors.
Note, however, that the dispersions for each Hubble type are very large. For
some Sab galaxies the pitch angles can be larger than for some Scd galaxies.
So, it is obvious that the weight of tightness for spiral patterns in the RC3
catalogue (de Vaucouleurs et al., 1991)
is only qualitative, not quantitative. 

\subsection{HI linewidths vs. pitch angle}

\begin{figure}
\resizebox{\hsize}{!}{\rotatebox{-90}{\includegraphics{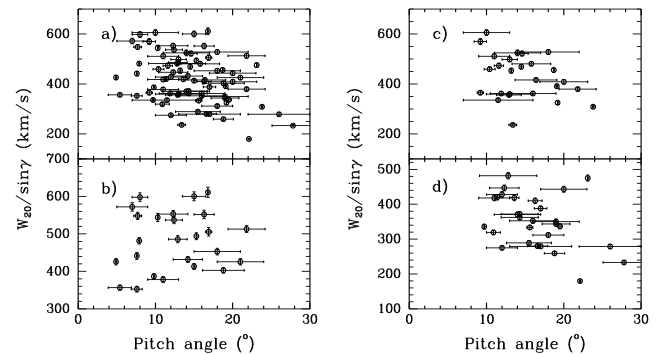}}}
%\vspace{-1.0cm}
\caption{HI linewidths plotted against measured pitch angle,
a) all Hubble types, b) Sab and Sb, c) Sbc, d) Sc and  Scd}
\label{FigVibStab}
\end{figure}

Tully \& Fisher (1977) studied the correlation
between distance independent
observable, global neutral hydrogen line profile
widths and absolute magnitude or diameter of spiral galaxies.
Using these measures, they offered a tool of measuring extragalactic distance, as well as
a method of estimating the Hubble constant. In this study, we will
present some statistical properties of spiral arms with some
physical properties of galaxies using
the weighted mean HI profile linewidths defined at the level of
$20\%$ of the peak flux (called W$_{20}$) from the RC3
catalogue (de Vaucouleurs et al., 1991).
These values are corrected for bandwidth, but
not for redshift.
The inclinations are derived by the Tully's formula (1988)
\begin{equation}
\gamma=\arccos\sqrt{[(\frac{d_{25}}{D_{25}})^{2}-0.2^2]/(1-0.2^2)}+
3^{\circ},
\end{equation}
here $D_{25}$ and $d_{25}$, which are the apparent major and minor isophotal
diameters measured at, or reduced down to the surface brightness level
$\mu_{B}=25.0$ B magnitudes per square
arcsecond, come from the RC3 catalogue (de Vaucouleurs et al., 1991).
The inclinations ($\gamma$) of all
these galaxies exceed $50^{\circ}$
from face-on so that there is no appreciable error in correcting
the hydrogen profile for projection. In Fig. 6, we plot the correlation
between pitch angle and W$_{20}$. The pitch angles
are from Danver (1942), Kennicutt (1981), Ma et al. (1997, 1998) and
other authors (Consid\`ere \& Athanassoula, 1988;
Puerari \& Dottori, 1992; Garc\'{\i}a-G\'omez \& Athanassoula,
1993). The horizontal error bars are from these sources,
or the r.m.s. calculated
from all values in the literature if there are different values
of the pitch angle for some particular galaxy. In this figure,
we do not include the galaxies for which the values of the r.m.s. of the pitch
angles are larger than 5 degrees.
The panel a) shows the correlation pooling
together all the data, b) the data for Sab and Sb galaxies,
c) the data for Sbc galaxies and d) the data for Sc and Scd galaxies.
The panel a) shows that there exits a correlation
between arm shape and W$_{20}$, which has been presented by
Kennicutt \& Hodge (1982).
However, the data  show
a great dispersion. Besides, as HI linewidths and pitch
angles both are somewhat related through Hubble types,
this panel can be dominated by the effect of the Hubble
sequence. We may not conclude from this panel alone that
the pitch angles are related to the W$_{20}$.
In order to
eliminate the effect of the Hubble sequence, we show the
panels b), c) and d) separating the galaxies in their
respective Hubble types. For each group, we calculated the Spearman rank-order
correlation coefficient and its significance. Table 5 summarises
the relevant quantities. In Col. 1 we list the symbol of
the panel in Fig. 6 and in Col. 2 the number of data
used for each panel. Col. 3
gives the correlation coefficient and Col. 4
its significance.
From Fig. 6 and Table 5, we can see that, except for the case of
Sab-Sb galaxies,
there exists a correlation
between the arm shape and W$_{20}$ for each Hubble
type separately. The case of panel b) can be due in part
to the difficulties in determining the HI line widths
for these galaxies, due to the lower HI contents,
specially in the central parts.
However, it is clear that
W$_{20}$ cannot be the only parameter that dictates arm
shape, since the scatter is large.

As we know, more massive galaxies have more stars than less massive ones, and
more massive galaxies rotate faster. W$_{20}$ connects with the maximum
rotation velocity of a galaxy. Tully \& Fouqu${\rm \acute{e}}$
(1985) proposed a formula
that could transform the 20$\%$ linewidths into the parameter W$_R$, which
approximates twice the maximum rotation velocity of a galaxy. The
correlation between pitch angle and W$_{20}$ can indicate that,
the arm shape of spiral galaxies is partially determined by the mass of a galaxy.

\subsection{Total mass luminosity ratio and total mass surface density vs. pitch angle}

Roberts \& Haynes (1994) systematically studied the physical parameters
along the Hubble sequence. They showed
that the total mass luminosity ratios ($M_T/L_B$) of spiral galaxies
are nearly independent of Hubble type, and that the total mass surface
densities ($\sigma_T$) are also approximately constant along the Hubble
sequence for Sab to Sc galaxies, showing some decrease for Scd galaxies.
Fig. 7 shows the correlations between pitch angle
and total mass luminosity ratio (panel a), and total mass surface density
(panel b). The sources of pitch angles and horizontal error bars are the same as
in Fig. 6. The total mass luminosity ratios and total mass surface densities
are calculated as in Roberts \& Haynes (1994) using
the RC3 catalogue (de Vaucouleurs et al., 1991).
In this figure, the inclinations ($\gamma$) of all
these galaxies also exceed $50^{\circ}$
from face-on so that there is no appreciable error in correcting
the hydrogen profile for projection. For simplicity,
as in Roberts \& Haynes (1994),  we use the corrected 20$\%$ linewidths
as the measure of $2V_{rot}$ and $D_{25}$ as the
indicator of $2R$.
For each correlation, we calculated the Spearman rank-order
correlation coefficient and its significance. Table 6 summarises
some of the relevant quantities. In Col. 1 we list the symbol of
the corresponding panel in Fig. 7 and in Col. 2 the number of galaxies. Col. 3
gives the correlation coefficient and Col. 4
its significance.
From Fig. 7 and Table 6, we can see
a significant trend in the sense that for lower total mass luminosity ratio
and lower mass surface density the pitch angles tend to be larger,
although the scatter in this figure is also large.

\begin{figure}
\resizebox{\hsize}{!}{\rotatebox{-90}{\includegraphics{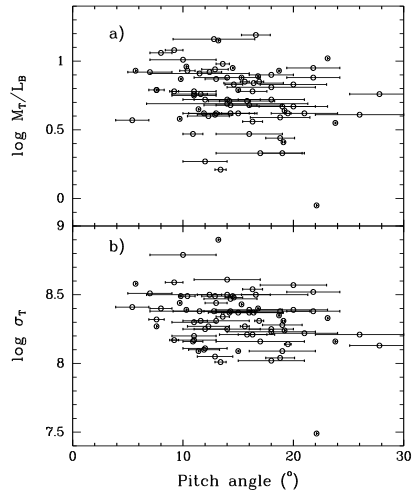}}}
\caption{Total mass luminosity ratio and total mass surface density
plotted against measured pitch angle.}
\label{FigVibStab}
\end{figure}

\section{Conclusion}

In this paper, we investigate some new statistical correlations between
the physical properties of disks and of spiral arms with some physical
properties of galaxies. Our main conclusions are:
(1). Along the Hubble sequence, the spiral galaxies tend to be thinner
and somewhat bluer, specially the later types, Sc-Scd.
(2). For later Hubble types, there is a correlation between arm pitch
angle and  HI linewidths. This correlation exits for each Hubble type
separately and can indicate that
the arm shape is partially determined by the mass of the galaxy.
(3). As the pitch angles increase, the total mass luminosity ratios and
total mass surface densities decrease.

\begin{acknowledgements}
We are indebted to the referee, Dr. Carlos Garc\'{\i}a-G\'omez,
for many critical comments and helpful suggestions, and
for English editing that
have greatly improved our paper.
\end{acknowledgements}

\newpage
\begin{table*}
\caption[]{Parameters of the distribution of thicknesses in the different Hubble type}
\begin{tabular}{cccccc}
\hline
\hline
Hubble type &       & Number of points & Mean & Dispersion & t-Std.\\
\hline
Sab and Sb  &            & 18  & 1.53  &  0.19 &       \\
Sbc         &            & 24  & 1.23  &  0.15 & 0.05  \\
            & Difference &  6  & 0.30  &  0.04 &       \\
\hline
Sab and Sb  &            & 18  & 1.53  &  0.19 &       \\
Sc          &            & 21  & 0.87  &  0.10 & 0.20  \\
            & Difference &  3  & 0.66  &  0.09 &       \\
\hline
Sbc         &            & 24  & 1.23  &  0.15 &       \\
Sc          &            & 21  & 0.87  &  0.10 & 0.07  \\
            & Difference & 3   & 0.36  &  0.05 &       \\
\hline
Sc          &            & 21  & 0.87  &  0.10 &       \\
Scd         &            & 4   & 0.20  &  0.02 & 0.05  \\
            & Difference & 17  & 0.67  &  0.08 &       \\
\hline
\end{tabular}
\end{table*}

\begin{table*}
\caption[]{Parameters of the distribution of $\overline{\log R_0}$ in the different Hubble type}
\begin{tabular}{cccccc}
\hline
\hline
Hubble type &   & Number of points & Mean & Dispersion & t-Std.\\
\hline
Sab and Sb  &            & 20 & 0.75  &  0.04 &       \\
Sbc         &            & 24 & 0.79  &  0.04 &  0.25 \\
            & Difference & 4  & 0.04  &  0.00 &       \\
\hline
Sbc         &            & 24 & 0.79  &  0.04 &       \\
Sc          &            & 22 & 0.90  &  0.04 &  0.07 \\
            & Difference & 2  & 0.11  &  0.00 &       \\
\hline
Sc          &            & 22 & 0.90  &  0.04 &       \\
Scd         &            & 6  & 1.20  &  0.15 &  0.06 \\
            & Difference & 16 & 0.30  &  0.11 &       \\
\hline
\end{tabular}
\end{table*}

\begin{table*}
\caption[]{Parameters of the distribution of $B-V$ color in the different Hubble type}
\begin{tabular}{cccccc}
\hline
\hline
Hubble type &  & Number of points & Mean & Dispersion & t-Std.\\
\hline
Sab and Sb  &            & 18 & 0.65  & 0.03 &       \\
Sbc         &            & 24 & 0.59  & 0.01 &  0.61 \\
            & Difference &  6 & 0.06  & 0.02 &       \\
\hline
Sab and Sb  &            & 18 & 0.65  & 0.03 &       \\
Sc          &            & 21 & 0.53  & 0.02 &  0.92 \\
            & Difference &  3 & 0.12  & 0.01 &       \\
\hline           
Sbc         &            & 24 & 0.59  & 0.01 &       \\
Sc          &            & 21 & 0.53  & 0.02 &  0.76 \\
            & Difference & 3  & 0.06  & 0.01 &       \\
\hline
Sc          &            & 21 & 0.53  & 0.02 &       \\
Scd         &            & 4  & 0.47  & 0.04 &  0.13 \\
            & Difference & 17 & 0.06  & 0.02 &       \\
\hline
\end{tabular}
\end{table*} 

\begin{table*}
\caption[]{Linear correlation coefficients}
\begin{tabular}{ccc}
\hline
\hline
Fig. 4      & Number of points & Correlation coefficient\\
\hline
a)  & 35    & 0.71  \\
b)  & 21    & 0.74  \\
c)  & 20    & 0.63  \\
d)  & 13    & 0.69  \\
\hline
\end{tabular}
\end{table*}

\begin{table*}
\caption[]{Spearman rank-order correlation coefficients}
\begin{tabular}{cccc}
\hline
\hline
Fig. 6      & Number of points & Correlation coefficient & Significance\\
\hline
a)  & 82    & -0.34 & 0.0016 \\
b)  & 25    &  0.15 & 0.48   \\
c)  & 27    & -0.35 & 0.071  \\
d)  & 30    & -0.40 & 0.029  \\
\hline
\end{tabular}
\end{table*}

\begin{table*}
\caption[]{Spearman rank-order correlation coefficients}
\begin{tabular}{cccc}
\hline
\hline
Fig. 7      & Number of points & Correlation coefficient & Significance\\
\hline
a)  & 82   & -0.22 & 0.050 \\
b)  & 82   & -0.24 & 0.038 \\
\hline
\end{tabular}
\end{table*}
\end{document}